\newcommand{\version}{November 17, 2000}

\documentclass[12pt]{article}
\usepackage{a4,amsthm,amsfonts,latexsym}

\swapnumbers
\theoremstyle{plain}
\newtheorem{thm}{THEOREM}
\newtheorem{cl}[thm]{COROLLARY}
\newtheorem{lem}[thm]{LEMMA}
\newtheorem{proposition}[thm]{PROPOSITION}
\theoremstyle{definition}
\newtheorem{rem}[thm]{Remark}

\newcommand{\beq}{\begin{equation}}
\newcommand{\eeq}{\end{equation}}
\def\beqa{\begin{eqnarray}}
\def\eeqa{\end{eqnarray}}

\newcommand{\R}{{\mathbb R}}

\newcommand{\Z}{{\mathbb Z}}

\newcommand{\Ll}{{\mathcal L}}
\newcommand{\Hh}{{\mathcal H}}
\newcommand{\G}{{\mathcal G}}
\newcommand{\Ginfty}{{\mathcal G}_\infty}
\newcommand{\Cc}{{\mathcal C}}
\newcommand{\eps}{\varepsilon}
\newcommand{\dx}{\frac\partial{\partial x}}
\newcommand{\dy}{\frac\partial{\partial y}}
\newcommand{\infspec}{{\rm inf\ spec\ }}
\newcommand{\half}{\mbox{$\frac{1}{2}$}}

\newcommand{\rperp}{r_\perp}
\newcommand{\x}{{\bf x}}
\newcommand{\A}{{\bf a}}

\date{\small\version}

\begin{document}
\markboth{\scriptsize{BS \version}}{\scriptsize{BS \version}}

\title{\bf{On the ordering of energy levels in homogeneous magnetic fields}}
\author{\vspace{5pt} Bernhard Baumgartner$^1$ and Robert Seiringer$^2$\\
\vspace{-4pt}\small{Institut f\"ur Theoretische Physik, Universit\"at Wien}\\
\small{Boltzmanngasse 5, A-1090 Vienna, Austria}}

\maketitle

\begin{abstract}
We study the energy levels of a single particle 
in a homogeneous magnetic field and in an axially symmetric external potential. 
For potentials that are superharmonic off the central axis,
we find a general ``pseudoconcave'' ordering of the ground state energies of the Hamiltonian restricted to
the sectors with fixed angular momentum.
The physical applications include 
atoms and ions in strong magnetic fields. There the energies are monotone 
increasing and concave in angular momentum. 
In the case of a periodic chain of atoms the pseudoconcavity 
extends to the entire lowest band of Bloch functions.  
\end{abstract}

\footnotetext[1]{E-Mail:
\texttt{baumgart@ap.univie.ac.at}}
\footnotetext[2]{E-Mail:
\texttt{rseiring@ap.univie.ac.at}}

\section{Introduction}

We consider the non-relativistic quantum mechanical theory of a single 
particle in a homogeneous magnetic field and in an external potential. 
The study of such systems is currently of interest in the context of 
theories of atoms in strong magnetic fields. 
In the way of describing the atoms as an assembly of electrons in an 
effective potential, an essential question is where the electrons are located. 
In the case of a very strong magnetic field this problem is connected with the 
comparison of individual energy levels, as the atom is not to be described 
with a completely semiclassical theory.
See \cite{LSY94,BSY00,HS00}. 
In all these studies the superharmonicity of the potential off the nucleus 
turned out to be the right property to deduce theorems on the 
localization of the electrons and on level ordering. 
It is precisely this property, superharmonicity off a certain axis, 
which is also used in our extension of the present knowledge 
presented in this paper. 

The potential is assumed to be invariant under rotations around the $z$-axis, 
the magnetic field in $z$-direction, so that the eigenvalue $m$ of $L_z$, 
the $z$-component of the angular momentum, is a ``good quantum number''. 
We may ignore spin and intrinsic magnetic moment, since this degree of 
freedom decouples in this case from the spatial behavior. Assuming appropriate 
gauging and appropriate units, either with a positive charge of the particle 
and the magnetic field pointing into the positive $z$-direction, or with a 
negative charge of the particle 
and the magnetic field pointing into the negative $z$-direction, 
the Hamiltonian acts in $\Ll^2(\R^3,d^3\x)$ as
\beq\label{ham}
H=-\Delta - B\,L_z + \frac{B^2}4 \rperp ^2 + V(\rperp,z),
\eeq
where $B>0$ denotes the absolute value of the strength of the magnetic field, 
$\rperp = \sqrt{x^2+y^2}$, $\x=(x,y,z)$. 
See \cite{AHS78} for details on the definition of $H$.

The Hamiltonian can be considered as a direct sum of $H_m$, where each $H_m$ acts 
in the subspace of eigenfunctions of $L_z$ with $L_z=m$. We compare the spectra 
of the operators $H_m$, denoting 
\beq
E_m = \infspec H_m.
\eeq 
Since each $H_{-m}$ is unitarily equivalent to $H_m + 2mB$, we can restrict the 
investigation to non-negative $m$. 
(Note that, for $V=0$, $E_m=B$ for all $m\geq 0$.)

If $V(\rperp,z) = V_\perp (\rperp) + V_z(z)$, the three-dimensional system 
splits into a one-dimensional $z$-dependent and a two-dimensional
$\rperp$-dependent system, and all the level comparison theorems concerning
comparison with a two-dimensional oscillator can be applied, with obvious
modifications. See \cite{BGM85}, the Subsections 4.7 and 6.4 of \cite{B85}, and
\cite{B91} for a simpler proof; see also Theorem B in \cite{GS95}.

But here we are interested in general cases without a splitting 
into $\rperp$-depen\-dent and $z$-dependent systems.
What is known up to now in these general situations is:

\begin{enumerate}
\item[(i)] There are examples, where $\inf\{E_m\}$ is attained as a minimum, 
  but not at $m=0$, \cite{LC77,AHS78}.
\item[(ii)] For all $m\geq 0$ we have $E_m\leq E_{m+1}+B$, with strict inequality 
if the $E_m$'s are eigenvalues.
\item[(iii)] If $E_m$ and $E_{m+1}$ are eigenvalues, $V(\rperp ,z)$ not constant in $\rperp $, 
  then $\partial V /\partial \rperp \geq 0$ implies $E_{m+1} > E_m$, 
  and $\partial V /\partial \rperp  \leq 0$ implies $E_{m+1} < E_m$, as shown by Grosse and Stubbe in 
  \cite{GS95}.
\item[(iv)] If $\infspec H$ is an eigenvalue, $\Delta V \leq 0$ (in the sense 
  of distributions) and $\Delta V \not= 0$, both at $\rperp>0$, the corresponding 
  eigenvector is unique and has angular momentum $m=0$, \cite{BS00}.
\end{enumerate}

The general bound (ii), which holds also for higher eigenvalues, is shown in Lemma \ref{gbd} in the Appendix, 
where also a short version of the proof of (iii) is given.

\begin{rem}[On proving (iv)]
In \cite{BS00} we considered $V = -\frac 1{|\x|}+\frac 1{|\x|}*\rho$, 
with $\rho$ a nice axially symmetric positive function. 
But the proof can be extended, without any change, to the more general case, 
where negative electric charges are located on the $z$-axis, and where 
$\rho$ is any repulsive axially symmetric distribution 
which is not zero everywhere away from the $z$-axis. 
Actually, this theorem 
will be a corollary of the central mathematical result of this paper, 
Theorem \ref{pseudoconcave} (at least if
$\infspec H$ is a discrete eigenvalue).
\end{rem}

Posing conditions on the potential $V$ in Section \ref{extern}, 
we try to allow for a wide variety 
of physical applications.
One of special importance is dealing with the energies of a 
charged particle in the electric field generated by attractive charges
situated on the $z$-axis and by an axially symmetric repulsive charge-distribution. 
It is  

\begin{thm}[Monotonicity and concavity for finite charges]\label{finite}
Let
\beq \label{vfinite}
V(\rperp,z) = -\int_{-\infty}^\infty \frac{1}{|\x-(0,0,z')|}\sigma (z')dz' 
+ \int_{\R ^3}\frac{1}{|\x-\x'|} \rho (\rperp ',z')d^3\x ',
\eeq
with $\sigma (z)dz$ a positive finite Borel measure on $\R$, and
$\rho (\rperp,z)d^3\x$ a non-negative finite Borel measure on $\R ^3$.
In these cases the sequence $E_m$ is non-decreasing, concave, and
$\lim_{m\to\infty} E_m = B$. Moreover, if $\int \rho (\rperp,z)d^3\x < \int
\sigma (z)dz$, the sequence $E_m$ is strictly monotone increasing and strictly
concave. \end{thm}
The proof will be given at the end of Section \ref{extern}.

Another physical application deals with the entire lowest band of 
energies in an infinite periodic system. It will be stated in
Theorem \ref{bloch}.

\section{Ordering in external potentials}\label{extern}

Consider the ``annihilation operator''
\beq
a=\frac 12\left(\sqrt\frac2B
\left(\dx-i\dy\right)+\sqrt\frac{B}2\left(x-iy\right)\right),
\eeq
and its adjoint, the ``creation operator'' $a^\dagger$.
These operators obey the commutation relation
\beq\label{comm}
[a,a^\dagger]=1,
\eeq
and the operator $a$ lowers the angular momentum,
\beq
[L_z,a]=-a.
\eeq
In absence of external potentials both $a$ and $a^\dagger$ commute with the Hamiltonian 
\beq
H_B = -\Delta - B\,L_z + \frac{B^2}4 \rperp ^2.
\eeq
Given any multiplication 
operator $V$, a straightforward calculation shows the formal equation
\beq\label{formcomm}
[a^\dagger,[a,V]]=[[V,a^\dagger],a]=\frac 1{2B}\left(-\frac{\partial^2
V}{\partial x^2}-\frac{\partial^2 V}{\partial y^2}\right).
\eeq
It holds in the sense of distributions, if 
$V \in \Ll^1_{\rm loc}$, and this condition will be fulfilled as
a consequence of the following
\medskip

\noindent{\bf Technical condition on $V$}:
\smallskip

\noindent
For fixed angular momentum $L_z=m$, $V$ is relatively bounded with
respect to $H_B$, with some bound less than $1$; i.e., for some $b_1<1$ and $b_2$,
depending on $m$,
\beq\label{relbound}
\|V\psi\|\leq b_1\|H_B\psi\|+b_2\|\psi\|
\eeq
for all $\psi$ with $L_z\psi=m\psi$. Note that this condition implies in
particular that $V\in\Ll^2_{\rm loc}$.

\medskip
The condition (\ref{relbound}) guarantees that each $H_m$ is semibounded and
essentially self-adjoint on $\Cc_0^\infty$, and $H$ is well defined by
$H=\bigoplus_m H_m$, without having to assume the semiboundedness of $H$.

The following proposition provides a large class of examples of 
potentials that fulfill the technical condition
stated above, and is applicable for most practical purposes.

\begin{proposition}\label{blochprop}
Assume that $V(\x)$ is locally $\Ll^2$, and there exists a constant,
which we denote as $\|V\|_{2,{\rm loc}}$, such that for all $n$ and some
$R>0$
\beq\label{condi}
\int_n^{n+1} dz \int_{\rperp<R} dx\,dy\, |V(\x)|^2 \, \leq \|V\|^2_{2,{\rm
loc}}.
\eeq
Moreover, we assume that there exist constants $C$ and $C_V<B^2/4$, such that
\beq\label{condii}
|V(\rperp,z)| \leq C+C_V\,\rperp^2\quad {\rm for\ all}\quad \rperp>R.
\eeq
Then $V$ fulfills the condition (\ref{relbound}).
\end{proposition}

\begin{proof}
Note that for operators $A\geq 0$ and $D^*=D$ the inequality
\beqa\nonumber
(A+D^2)^2&=&A^2+D^4+2DAD-\left[[D,A],D\right]\\
&\geq& A^2+D^4- \left[[D,A],D\right]
\eeqa
holds. Using this with $A=-\Delta$ and $D=B\rperp/2$ we get, for any $\psi$,
\beq\label{deltahb}
\|\Delta\psi\|^2+\left(\frac B2\right)^4\|\rperp^2\psi\|^2\leq
\|\left(-\Delta+\frac{B^2}4 \rperp^2\right)\psi\|^2+B^2\|\psi\|^2.
\eeq
Denoting $V_1(\x)=V(\x)\theta(R-\rperp)$ and $V_2=V-V_1$, we can estimate
\beqa\nonumber
\|V\psi\|&\leq& \|V_1\psi\|+\|V_2\psi\|\\
&\leq& \|V_1\psi\|+\frac{4C_V}{B^2}\|H_B\psi\|+\left( \frac{4C_V}{B}(|m|+1) + C
\right)\|\psi\|,
\eeqa
where we used $L_z\psi=m\psi$. Now $V_1$ is relatively $-\Delta$-bounded with
arbitrary small bound (\cite{RS78}, Thm. XIII.96), and by (\ref{deltahb}) the
same holds with $-\Delta$ replaced by $H_B$.
\end{proof}

For $m\geq 0$ let now $\psi_m$ denote the ground state wave
function of $H_m$, if it exists. Writing
\beq
\psi_m(\x)=e^{im\varphi}\rperp^m f(\rperp,z),
\eeq
where $(\rperp,\varphi)$ denote polar coordinates for $(x,y)$, we see that $f$
is a ground state for
\beq
\widetilde H_m=-\frac{\partial^2}{\partial \rperp ^2}-\frac{2m+1}\rperp
\frac{\partial}{\partial \rperp}-\frac{\partial^2}{\partial
z^2}+\frac{B^2}4 \rperp^2-m\,B+V(\rperp,z)
\eeq
on $\Ll^2(\R^3,\rperp^{2m}d^3\x)$. If $\widetilde H_m$ has a ground state, it
is unique and strictly positive. So $\psi_m$ is unique. Moreover, $f$ is a
bounded H\"older continuous function, so $\psi_m$ behaves like $\rperp^m$
for small $\rperp$; i.e. $|\psi _m(\x)| \leq C|\rperp|^m$.

We now have the necessary prerequisites to prove

\begin{thm}[Pseudoconcavity in angular momentum]\label{pseudoconcave}
If $V(\rperp,z)$ fulfills the technical condition stated above, and if 
the Laplacian of the potential off the z-axis is non-positive, 
the ground state energies $E_m$ of the restricted Hamiltonians $H_m$ 
obey the following inequalities, if $m \geq 1$: 
\beq\label{conclude0}
E_m \geq \min\{E_{m-1},E_{m+1}\},
\eeq
\beq\label{concave}
E_m \geq \frac12 (E_{m-1}+E_{m+1}) \qquad {\rm if} \quad E_m\geq E_{m-1},
\eeq
\beq\label{psconcave}
E_m \geq \frac1{2m+1} (mE_{m-1}+(m+1)E_{m+1}) \qquad {\rm if} \quad E_m\geq E_{m+1}.
\eeq
If $\Delta V$ is not vanishing everywhere off the z-axis, 
and if $E_m$ is a discrete eigenvalue, 
the inequalities (\ref{conclude0}), (\ref{concave}) and (\ref{psconcave}) are strict. 
\end{thm} 

\begin{proof}
We add $H_B-E_m$ to $V$ in (\ref{formcomm}), use $[H_B,a]=0$,
note that $H=H_B+V$, take the
expectation value with $\varphi \in \Cc_0^\infty$ and expand the double commutator.
We get, using moreover $\partial^2 /\partial x^2 + \partial^2 /\partial y^2 = 
\Delta -\partial^2 /\partial z^2$, 
\beqa\label{eq1}
\langle \varphi|a^\dagger (H-E_m) a+a (H-E_m) a^\dagger|\varphi\rangle
=\frac1{2B}\langle\varphi|\Delta V|\varphi\rangle 
-\frac1{2B}\langle\varphi|\frac{\partial^2V}{\partial z^2}|\varphi\rangle \nonumber \\
+\langle\varphi|a^\dagger a(H-E_m)
+(H-E_m)aa^\dagger |\varphi\rangle.
\eeqa
Now assume that $E_m$ is a discrete eigenvalue and let $\psi_m$ be the
corresponding ground state. Since $H$ is a closed operator and $\Cc_0^\infty$
is a core of $H$, there exist sequences of wave functions $\varphi_k \in
\Cc_0^\infty$ with $\|\varphi_k\|=1$ and angular momentum $m$, converging in
norm to $\psi_m$ such that also $H\varphi_k \to H\psi_m=E_m\psi_m$ in norm.
Since
\beq
2Ba^\dagger a=H_B+\frac{\partial^2 }{\partial z^2}+2BL_z-B ,
\eeq
condition (\ref{relbound}) on $V$ guarantees that
the operators $a^\dagger a$ and $aa^\dagger$ are bounded relative to $H_m$
on $\Hh_m$, the subspace of $\Hh$ where $L_z=m$.
It follows that, as $k\to\infty$, also $a^\dagger a\varphi_k$ and $aa^\dagger\varphi_k$
converge and are bounded in norm, and the second line in (\ref{eq1}),
with $\varphi =\varphi_k$, converges to $0$.

Set $\A=(0,0,a)$. We claim that it is no restriction to assume that the
functions \beq
f_k(a)\equiv \int |\varphi_k(\x+\A)|^2 V(\x) d^3\x
\eeq
have their minimum at $a=0$. To see this, let
$\tilde\varphi_k(\x)=\varphi_k(\x-\A_k)$, where $\A_k=(0,0,a_k)$ is chosen such
that $f_k(a_k)\leq f_k(a)$ for all $a$. This is possible since by assumption
$E_m$ is a discrete eigenvalue, so one can not attain any 
expectation value in the gap above $E_m$ by shifting 
$\varphi_k$ to infinity. Now 
\beq
\langle\tilde\varphi_k|H|\tilde\varphi_k\rangle\leq
\langle\varphi_k|H|\varphi_k\rangle \longrightarrow E_m,
\eeq
so there exists a subsequence, again denoted by $\tilde\varphi_k$, such that
$\tilde\varphi_k\rightharpoonup \psi_m$ weakly in $\Ll^2$. Since also the norms
converge, there is even strong convergence. In particular, $a_k\to 0$ as
$k\to\infty$. Therefore, since $H_B$ is translation invariant in $z$-direction,
and since $V$ is relatively bounded,
\beq
\|H(\tilde\varphi_k-\psi_m)\|\longrightarrow 0.
\eeq
Now denote $\tilde\varphi_k$ by $\varphi_k$, which proves our claim.
With $\varphi_k$ chosen as above, we have
\beq\label{shift}
\int |\varphi_k|^2 \frac{\partial^2 V}{\partial z^2}=\left.
\frac{\partial^2}{\partial a^2} f_k(a)\right|_{a=0} \geq 0.
\eeq

Concerning $\langle \varphi_k |\Delta V |\varphi_k\rangle$,
the H\"older continuity of $\psi_m$ (\cite{LL97}, Theorem 11.7)
and the fact that $\psi_m(\rperp =0,z)=0$ if $m \geq 1$ 
allows us to integrate $|\psi_m|^2\Delta V$
and apply Fatou's Lemma to conclude that
\beq\label{deltapart}
\limsup_{k\to\infty} \frac1{2B}\langle\varphi_k|
\Delta V|\varphi_k\rangle \leq 
\frac1{2B}\langle\psi_m|\Delta V |\psi_m\rangle,
\eeq
where we used that $\varphi_k(\x)\to\psi_m(\x)$ pointwise. (The convergence is
even uniform in $\x$, since $\|\Delta\varphi_k\|$ is uniformly bounded by
(\ref{deltahb}).)
If $\Delta V<0$ somewhere away from the $z$-axis,  
the fact that $|\psi_m|>0$
away from $\rperp=0$   
implies that the right hand side of (\ref{deltapart}) is
strictly negative.
Using this and (\ref{shift}) in
(\ref{eq1}) we can conclude that \beq\label{conclude}
\limsup_{k\to\infty}\langle\varphi_k|a^\dagger(H-E_m)a+a(H-E_m)a^\dagger|\varphi_k\rangle < 0.
\eeq

Now, since $a^\dagger|\varphi_k\rangle$ has angular momentum $L_z=m+1$, and 
$a|\varphi_k\rangle$ has angular momentum $L_z=m-1$, this implies 
that the spectrum of either $H_{m+1}$ or $H_{m-1}$ must reach below $E_m$, that is 
\beq\label{conclude2}
E_m > \min\{E_{m-1},E_{m+1}\} \qquad {\rm if} \quad m\geq 1.
\eeq

To prove the relations (\ref{concave}) and (\ref{psconcave}) 
in their strict form we use the commutation relation (\ref{comm}) 
and observe that
$\|a^\dagger\varphi_k\|^2=\|a\varphi_k\|^2+1$. 
So the inequality (\ref{conclude}) tells us moreover that 
\beq\label{limsup}
\limsup_{k\to\infty}\left( (E_{m-1}-E_m) \|a \varphi_k\|^2 +
(E_{m+1}-E_m)(\|a \varphi_k\|^2 +1)\right) < 0.
\eeq
In the case $E_m\geq E_{m-1}$, we note that the strict form of 
(\ref{concave}) holds trivially, if $E_m>E_{m+1}$. 
Otherwise, if $E_m\leq E_{m+1}$, (\ref{limsup}) gives 
\beq \label{concave0}
\limsup_{k\to\infty} \|a\varphi_k\|^2 (E_{m-1}+E_{m+1}-2 E_m) <
E_m-E_{m+1}\leq0.
\eeq
In the case $E_m\geq E_{m+1}$ we use $\|a\varphi_k\|^2\geq m>0$, 
which is a consequence of $a^\dagger a\geq L_z$. 
Using $1/m \geq 1/\|a \varphi_k\|^2$, the inequality (\ref{limsup}) gives 
\beq \label{psconcave0}
\limsup_{k\to\infty}\|a \varphi_k\|^2 
\left( (E_{m-1}-E_m) + (E_{m+1}-E_m)(1+\frac1m)\right)<0.
\eeq
So our assertions on the strict inequalities follow.

If $H_m$ does not have a discrete eigenvalue at the bottom of its spectrum, 
or if $\Delta V$ happens to be zero everywhere off the $z$-axis,
add $\varepsilon W$ to the potential, with the superharmonic function 
$W(\rperp,z)=(z^2 - \rperp^2)$, and $\eps$ small enough (more precisely,
$0 < \varepsilon < B^2(1-b_1)/4$, with $b_1$ as in (\ref{relbound})). 
The Hamiltonians $H_m(\eps)=H_m+\varepsilon W$ do have 
discrete ground states for each $m$ (\cite{RS78}, Theorem XIII.67), 
and (\ref{conclude2}), as well as (\ref{concave0}) and 
(\ref{psconcave0}) hold for 
the corresponding energies $E_m(\varepsilon)$. 
Strictly speaking, the potentials $V+\eps W$ are not relatively bounded
with respect to $H_B$ on $\Hh_m$, but it is not difficult to see, 
using boundedness relative to $H_B+\varepsilon z^2$ instead, that the
conclusions above remain valid. 
Now taking the limit $\varepsilon \to 0$ leads to the inequalities  
(\ref{conclude0}), (\ref{concave}) and (\ref{psconcave}).
\end{proof}
\begin{rem}[Change of the conditions]\label{change}
Theorem \ref{pseudoconcave} and the following corollaries hold also 
when the condition $\Delta V \leq 0$ off the $z$-axis is replaced by 
$\partial^2 V /\partial x^2 +\partial^2V /\partial y^2\leq 0$ 
off the $z$-axis. The proof is essentially the same as above, 
but without any involvement of $\partial^2V /\partial z^2$, and 
discreteness of the ground state is not needed. 
\end{rem}

Considering the entire sequence $E_m$, the relation (\ref{conclude0}) 
obviously 
implies that an increase of $E_m$ in $m$ can turn into 
a decrease, but not vice versa. We state this and the 
global forms of the other relations as 

\begin{cl}[Global pseudoconcavity]\label{pseudoconcave2}
Under the same conditions on $V$ as above, 
there is an $M$, possibly $0$ or $\infty$, such that the sequence $\{E_m\}$ 
is strictly increasing and concave 
for $0\leq m \leq M$ and non-increasing for $m \geq M$. 
Moreover, the ground state energies of two neighboring values of 
angular momentum, $\ell, \ell+1$, determine upper bounds on all of 
$\{E_m\}$ by a tangential sequence: For all $\ell\geq 0$ and $m\geq 0$, 
\beq\label{glob1}
E_m \leq E_\ell +(m-\ell)(E_{\ell +1}-E_\ell) 
\qquad {\rm if} \quad E_\ell\leq E_{\ell +1},
\eeq
\beq\label{glob2}
E_m \leq E_\ell +(S_m-S_\ell)(\ell +1)(E_{\ell +1}-E_\ell) 
\qquad {\rm if} \quad E_\ell\geq E_{\ell +1},
\eeq
where 
\beq
S_m =\sum_{\mu=1}^m \frac 1\mu.
\eeq
These inequalities are strict, if some $E_\mu$ for $m\leq \mu\leq\ell$ 
or $\ell+1\leq \mu\leq m$ is a discrete eigenvalue, if $m\not\in\{\ell,\ell+1\}$ 
and if the Laplacian of the potential is not identically zero off the $z$-axis. 
\end{cl}
\begin{proof}
For $E_m$ in the increasing 
part of the sequence $\{E_m\}$, the inequality (\ref{glob1}) is 
the standard statement that 
a tangent line lies above a concave sequence. Since the linear 
tangential sequence is increasing, it is trivial to extend this 
inequality to the $E_m$ in the decreasing part. 

To prove (\ref{glob2}), consider $\Delta_m=(E_m -E_{m-1})$ and 
observe that (\ref{psconcave}) can be written as a
monotonicity relation for $m\Delta_m$, 
\beq
(m+1)\Delta_{m+1}\leq m\Delta_m \qquad {\rm if} \quad E_m\geq E_{m+1}.
\eeq
Starting with $\Delta_{\ell+1}$, which is not positive, this 
means $\Delta_\mu \leq \frac{\ell+1}\mu \Delta_{\ell+1}$ if $\mu \geq \ell+1$, 
and $\Delta_\mu \geq \frac{\ell+1}\mu \Delta_{\ell+1}$ if $\mu\leq \ell$. 
The extension of the latter inequality to the increasing part of 
$\{E_\mu\}$ is trivial.
Summing these inequalities for $\Delta_\mu$, 
either  for $m+1 \leq \mu\leq\ell$ 
or $\ell+1 \leq \mu\leq m$, gives (\ref{glob2}). 
Moreover, any strict inequality for one of the $\Delta_\mu$ gives 
(\ref{glob2}) as a strict inequality.
\end{proof}

We remark that the asymptotic behavior of the decreasing tangential 
sequence, $S_m \sim \ln m$, is optimal. This is demonstrated on the 
example with the infinite charged tube in Section \ref{phys}.

An immediate consequence of this logarithmic divergence is 

\begin{cl}[Either no decrease or infinite decrease]\label{monotone}
The \newline 
sequence $\{E_m\}$ is either not decreasing at all 
or it is decreasing to $-\infty$ as 
$m\to\infty$.
\end{cl}

Moreover, we have the generalization of the theorem on the ground state
stated in \cite{BS00}.

\begin{cl}[The ground state has zero angular momentum]\label{groundzero}
If \newline $V$ fulfills the conditions as in Theorem \ref{pseudoconcave}, and if $\infspec H$ 
is a discrete eigenvalue,
the corresponding eigenvector is unique and has zero angular momentum.
\end{cl}

\begin{proof}
Corollaries \ref{pseudoconcave2} and \ref{monotone} imply that either $E_0<E_m$ or $E_0=E_m$ for all $m$. Since $\infspec H$ is a discrete eigenvalue by assumption, it cannot be infinitely degenerate, so our assertion is proved.
\end{proof}

\begin{rem}[The hydrogen atom]
Corollary \ref{groundzero} applies also to the hydrogen atom
in a constant magnetic field, and 
provides a simple proof of $L_z=0$
in the ground state; the third proof following \cite{AHS81} 
and \cite{GS95}. 
\end{rem}

We now have the tools to prove Theorem \ref{finite}.

\begin{proof}[Proof of Theorem \ref{finite}.]
The potential $V$ is relatively bounded with respect to $-\Delta$ 
with relative bound $0$ on the {\it entire} Hilbert space. This implies that $V$  
fulfills the technical condition which is necessary 
to apply Theorem \ref{pseudoconcave} and Corollary \ref{monotone}, 
and moreover that the Hamiltonian is bounded below. 
So the sequence $E_m$ is non-decreasing and concave, and $\lim E_m=B$. 
Since the asymptotic behavior of the potential, as $\rperp \to \infty$, 
is $V(\x) \sim -(Z-C)/|\x|$, where $Z=\int \sigma dz$, 
$C=\int \rho \, d^3\x$, it is easy, if $C<Z$, to construct trial functions 
for large $m$, such that the expectation value of $H_m$ 
is lower than the edge of its essential spectrum, which is at $B$. 
Such trial wave functions are of the form 
$\Phi_m(x,y)\eps\varphi (\eps z)$, where $\Phi_m$ is the wave function 
in $\Ll^2(\R^2)$ 
with $L_z=m$ in the lowest Landau band, and $\eps$ has to be small enough. 
So each $E_m$ is a discrete eigenvalue, and increase and concavity are strict.
\end{proof}

\section{Physical applications and extensions}\label{phys}

{\bf The mean field model of a positively charged ion}
\newline
Consider
\beq\label{ion}
V = -\frac1{|\x|} + \rho * \frac1{|\x|}
\eeq
with $\rho = \rho (\rperp,z) \geq 0$ and $C=\int \rho < 1$. 
We can apply Theorem \ref{finite}, 
and infer that the $E_m$'s are monotone increasing eigenvalues.
\bigskip
\newline
{\bf The mean field model of a negatively charged ion}
\newline
The potential is as in (\ref{ion}), but now with $C>1$. 
Since $V(\x) \sim (C-1)/|\x|$ as $\rperp \to \infty$, 
and since the angular momentum barrier shields the nucleus, 
there are no bound states for large $m$. 
The energies $E_m$ may be increasing up to some finite $M$, 
and will all be equal to $B$ for $m\geq M$. 
\bigskip
\newline
{\bf Attractive point charge in a repulsive homogeneously charged hollow tube}
\newline
Consider the potential 
\beq\label{hollow}
V = -\frac1{|\x|} - \tau \theta (\rperp-R)\ln (\rperp /R)
\eeq
where $\tau$ and $R$ are positive constants. 
The second part of $V$ is the 
``renormalized'' potential of an infinite hollow tube with radius $R$. 
It is unbounded below as $\rperp \to\infty$, and this implies 
$E_m\to -\infty$ as $m\to\infty$. On the other hand, the 
ground states with angular momentum $m$ are localized near $\rperp\sim\sqrt{2m/B}$, 
and will not really ``feel'' the hollow tube, if they are inside. 
The $E_m$ will increase like the corresponding states of the free hydrogen atom, 
at least up to $m\sim R^2B/2$. 
So there is an increase of ground state energies followed by a decrease. Moreover, 
since the $E_m$ are varying continuously as functions of the parameters $\tau$ and 
$R$, there will be cases, where the maximum of the $E_m$ is attained twice, at 
two neighboring angular momenta simultaneously. 
The asymptotics of $\{E_m\}$ as $m \to\infty$ is, to leading order, 
\beq
E_m \sim -\frac\tau 2 \ln m .
\eeq
\bigskip
\newline
{\bf An infinite periodic chain of atoms}
\newline
Let $\A$ be the vector $(0,0,a)$, defining the periodicity in $z$-direction. 
Consider
\beq
V(\rperp,z) = \sum_{n\in\Z ,\,{\rm ren}}\left( -\frac1{|\x -n\A|} + \rho * \frac1{|\x -n\A|}\right),
\eeq
where the axially symmetric positive charge density $\rho$ is localized in the elementary slice 
$0\leq z<a$, and $\int\rho$ is assumed to be finite.
If $\int\rho\not=1$, the infinite sum has to be coupled with a renormalization: 
Let $D=1-\int\rho$. 
\beq
\sum_{n\in\Z ,\,{\rm ren}}...\equiv\lim_{N\to\infty}\left(\sum_{-N}^N...+2D\ln N \right).
\eeq
(Convergence is guaranteed at least for $\rho$ with compact support.) 
This potential is e.g. of interest for the study of chains of atoms 
in strong magnetic fields within the DM theory of \cite{LSY94}, 
as will be discussed in \cite{JRY00}. 
Finiteness of $\int\rho$ guarantees that V can be split into the sum 
of two parts, one of them uniformly locally $\Ll^2$, 
the other diverging only logarithmically as $\rperp \to\infty$. 
So Proposition \ref{blochprop} applies and 
$V$ fulfills the technical condition. 

Since the Hamiltonian commutes with a group of discrete translations in $z$-direction, 
the Hilbert space can be decomposed into the direct integral of $\Hh _{\alpha}$, 
$0\leq\alpha <2\pi$,
such that the Hamiltonian $H$ is represented as a direct integral of $H_{\alpha}$, 
where the domain of definition of $H_{\alpha}$ contains the Bloch-functions, 
continuous functions with the property 
\beq
\psi (x,y,z+a)=e^{i\alpha}\psi (x,y,z), 
\eeq
normalized by $\|\psi\|^2=\int_{\R ^2}dx\,dy\int_0^a|\psi|^2dz$.  
The ground state energies $E_m$ are attained in the space $\Hh_0$, but we can 
extend our results to all of the lowest band, to the $E_m(\alpha)$, the 
ground state energies of $H_m$ restricted to $\Hh_\alpha$. 

\begin{thm}[Pseudoconcavity of Bloch function energies]\label{bloch}
Consider potentials which are symmetric under a group of discrete
$z$-translations, i.e. $V(\rperp,z+a)=V(\rperp,z)$ for some $a>0$, and fulfill
the conditions of Prop. \ref{blochprop}. Let $H_\alpha$ and $H_{\alpha,m}$ be
the restrictions of $H$ and $H_m$ onto the space of Bloch functions,
and denote $E_m(\alpha) = \infspec H_{\alpha,m}$. 
Theorem \ref{pseudoconcave} and the Corollaries \ref{pseudoconcave2}, 
\ref{monotone} and \ref{groundzero} hold 
also for each of the $E_m(\alpha)$ instead of the $E_m$. 
The pseudoconcavity is moreover strict if $\Delta V$ is not identically 
zero off the $z$-axis. 
\end{thm}
\begin{proof}
One can mimic the proof of Theorem \ref{pseudoconcave}.
All the operations we made there commute with the $z$-translation. 
Since each $H_{\alpha,m}$ has discrete spectrum, as we show in
Proposition \ref{purepoint} in the Appendix,
the inequalities (\ref{conclude0}) -- 
(\ref{psconcave}) hold in strict form if $\Delta V\not= 0$. 
\end{proof}

\section*{Appendix}

\begin{lem}\label{sobolev}
Let $\Delta$ be the Laplacian with Neumann boundary conditions on
\beq\label{gn}
\G_n = \{\x , n\leq z < n+1\}.
\eeq
Then, for $p>3/2$ and for some constant depending on $p$,
\beq
-\Delta+V(\x)\geq -1-C(p) \|V_-\|_p^{2p/(2p-3)},
\eeq
where $V_-=min\{V,0\}$, and the norm is
for the restriction of $V_-$ onto $\G_n$.
\end{lem}

\begin{proof}
By the Sobolev inequality on $\G_n$ there is a constant $C>0$ such
that
\beq
\|\nabla f\|_2^2 + \|f\|_2^2 \geq C \|f\|_6^2
\eeq
for all $f\in \Hh^1(\G_n)$ (cf. \cite{A75}, Thm. 5.4). Therefore we can
estimate, for $\|\psi\|_2=1$,
\beq\label{psi6}
\langle\psi|-\Delta+V|\psi\rangle\geq -
1+C(R)\|\psi\|_6^2- \|V_-\|_p \|\psi\|_6^{3/p},
\eeq
where we used H\"olders inequality, with $p^{-1}+q^{-1}=1$, twice:
\beqa
|\langle\psi|V_-|\psi\rangle|&\leq &\|V_-\|_p \|\psi^2\|_q,\\
\|\psi^2\|_q &\leq &\|\psi\|_6^{3/p} \|\psi\|_2^{2-3/p}.
\eeqa
Optimizing (\ref{psi6}) with
respect to $\|\psi\|_6$ yields the desired result.
\end{proof}

\begin{proposition}[Discrete spectrum]\label{purepoint}
Each operator $H_{\alpha,m}$ defined in Theorem \ref{bloch} has discrete spectrum.
\end{proposition}

\begin{proof}
We split the potential $V$ into 
\beqa
V_{\rm loc} &=& \theta(R-\rperp )V \nonumber \\
{\rm and} \qquad V_\infty &=& \theta (\rperp -R)V. \nonumber
\eeqa
Then we split the Hamiltonian $H$ into 
\beqa
H_{\rm loc} &=& \delta H_B + V_{\rm loc} \nonumber \\
{\rm and} \qquad H_\infty &=& (1-\delta)H_B+V_\infty, \nonumber
\eeqa
where $\delta= \half (1-4C_V /B^2)$, $C_V$ as defined in Prop. \ref{blochprop}.
We do the same splittings for all the $H_{\alpha,m}$. 
We use that 
$\infspec H_{\alpha,m,{\rm loc}} \geq \infspec H_{N,m,{\rm loc}}$, 
where $N$ denotes the Neumann boundary conditions, 
use Lemma \ref{sobolev} and the condition (\ref{condi}) to find a lower bound
to $H_{\alpha,m,{\rm loc}}$ by a constant operator,
\beq
H_{\alpha,m,{\rm loc}} \geq -{\rm const.}(1+\|V\|_{2,{\rm loc}}^4) \equiv
-C(V). \eeq
Now we add $H_{\alpha,m,{\rm loc}}$ and $H_{\alpha,m,\infty}$ and use the
condition (\ref{condii}): 
\beq
H_{\alpha,m} \geq -C(V)-(1-\delta)(\Delta+Bm)-C+\frac12(\frac{B^2}4-C_V)\rperp ^2.
\eeq
The operator on the right side obviously has discrete spectrum, and  
the min-max principle implies that the same is true for $H_{\alpha,m}$. 
\end{proof}

We add a short version of the proof of the Grosse-Stubbe inequality. 
We need the following

\begin{lem}[General bound on the decrease]\label{gbd}
Let $E_{m,n}$ be the $n$'th eigenvalue (counted from below) of 
$H_m$, or the edge of its essential spectrum, if there are less than 
$n$ discrete eigenvalues. Then, if $m\geq 0$,
\beq\label{gbdi}
E_{m,n} \leq E_{m+1,n}+B.
\eeq
\end{lem}
\begin{proof}
We map each subspace $\Hh_m$ unitarily onto 
$\Ll^2(\R_+\times\R, 2\pi \rperp d\rperp \,dz )$, by writing, for 
$\psi\in\Hh_m$, 
\beq
\psi(\x)=e^{im\varphi}\chi (\rperp,z), 
\eeq
and mapping $\psi\to\chi$. 

The Hamiltonian $H_m$ is then unitarily equivalent to 
\beq
\hat H_m = -\frac{\partial^2}{\partial \rperp ^2}-\frac1\rperp
\frac{\partial}{\partial \rperp}-\frac{\partial^2}{\partial
z^2}+\frac{B^2}4 \rperp^2 +\frac{m^2}{\rperp^2} -m\,B+V(\rperp,z),
\eeq
and the right side of (\ref{gbdi}) is the $n$'th eigenvalue of 
\beq
\hat H_{m+1}+B=\hat H_m+\frac{2m+1}{\rperp^2}.
\eeq
This means that $\hat H_{m+1}+B \geq \hat H_m$, so (\ref{gbdi}) follows with 
the help of the min-max principle.
\end{proof}
\begin{proof}[Proof of the Grosse-Stubbe inequality]
The following formula can be interpreted as calculating the acceleration of 
the particle, subtracting the effect of the Lorentz force, and then taking 
a linear combination of the $x$- and $y$-components, denoted as $x_+=x+iy$:
\beq \label{accel}
[H,[H,x_+]]+2B[H,x_+]=2\frac {x_+} \rperp \frac{\partial V}{\partial\rperp}.
\eeq
Now we take the matrix elements between $\psi_{m+1}$ and $\psi_m$,
\beq
\left( (E_{m+1}-E_m)^2 +2B(E_{m+1}-E_m)\right)\langle\psi_{m+1}|x_+|\psi_m\rangle=
2\langle\psi_{m+1}|\frac {x_+} \rperp \frac{\partial V}{\partial\rperp}|\psi_m\rangle ,
\eeq
and choose the phases appropriately, so that we can write 
$\psi_m=(x_+/\rperp)^m\chi_m$ with $\chi_m\geq 0$. We get 
\beq
(E_{m+1}-E_m)(E_{m+1}-E_m+2B)\langle\chi_{m+1}|\rperp|\chi_m\rangle=
2\langle\chi_{m+1}| \frac{\partial V}{\partial\rperp}|\chi_m\rangle .
\eeq
The matrix element on the left side 
is positive, and by Lemma \ref{gbd} we know that $E_{m+1}-E_m+2B \geq B>0$. 
So the sign of $\frac{\partial V}{\partial\rperp}$, which determines the sign of the 
right side, determines also the sign of $(E_{m+1}-E_m)$ on the left side.
\end{proof}

The Grosse-Stubbe inequality can be extended in a weakened form to cases where the $E_m$ are 
not necessarily eigenvalues. To see this, we approximate $V(\rperp ,z)$ by 
$V_R(\rperp ,z) \equiv V(\min\{\rperp ,R\} ,z) + z^2/R$. With this potential the ground state 
energies $E_{R,m}$ 
have to be eigenvalues, and are strictly ordered. Then we consider 
the limit $R \to \infty$. Since $E_{R,m}\to E_m$, it follows that the 
monotonicities of $V$ in $\rperp$ imply monotonicities of the $E_m$ in $m$, 
which don't have to be strict if they are not eigenvalues.


\begin{thebibliography}{BGM85}


\bibitem[LSY94]{LSY94}  Elliott H. Lieb, Jan Philip Solovej,
and Jakob Yngvason:
{\it Asymptotics of Heavy Atoms in High Magnetic Fields:
I. Lowest Landau Band Regions},
Commun. Pure Appl. Math. {\bf XLVII}, 513--591 (1994)

\bibitem[BSY00]{BSY00}  Bernhard Baumgartner, Jan Philip Solovej, and Jakob
Yngvason:
{\it Atoms in strong magnetic fields:
The high field limit at fixed nuclear charge},
Commun. Math. Phys. {\bf 212}, 703--724 (2000)

\bibitem[HS00]{HS00}  Christian Hainzl, Robert Seiringer: 
{\it A discrete density matrix theory for atoms in strong magnetic fields},
arXiv:math-ph/0010005, to appear in Commun. Math. Phys.

\bibitem[AHS78]{AHS78} J. Avron, I. Herbst, B. Simon, {\it Schr\"odinger
Operators with Magnetic Fields. I. General Interaction},
Duke Math. J. {\bf 45}, 847-883 (1978)

\bibitem[BGM85]{BGM85} Bernhard Baumgartner, Harald Grosse, Andr\'e Martin, 
{\it Order of Levels in Potential Models},
Nucl. Phys. B{\bf 254}, 528-42 (1985)

\bibitem[B85]{B85} Bernhard Baumgartner, 
{\it Level Comparison Theorems},
Ann. Phys. (NY) {\bf 168}, 484-526 (1985)

\bibitem[B91]{B91} Bernhard Baumgartner, 
{\it Perturbation of Supersymmetric Systems in Quantum Mechanics}, in 
{\it Recent Developments in Quantum Mechanics},
A. Boutet de Monvel et al. (eds.), pp195-208,
Kluwer Acad. Publ., Netherlands, 1991

\bibitem[GS95]{GS95} Harald Grosse, Joachim Stubbe, {\it Splitting of Landau
Levels in the Presence of External Potentials},
Lett. Math. Phys. {\bf 34}, 59-68 (1995)

\bibitem[LC77]{LC77} Richard Lavine, Michael O'Carroll, {\it Ground state
properties and lower bounds for energy levels of a particle in a uniform 
magnetic field and external potential},
J. Math. Phys. {\bf 18}, 1908-12 (1977)

\bibitem[BS00]{BS00} Bernhard Baumgartner, Robert Seiringer, {\it Atoms with 
bosonic ``electrons'' in strong magnetic fields},
arXiv:math-ph/0007007, to be published in Annales Henri Poincar\'e.

\bibitem[AHS81]{AHS81}  J.E. Avron, Ira W. Herbst, and Barry Simon, {\it
Schr\"odinger Operators with Magnetic Fields III.
Atoms in Homogeneous Magnetic Field},
Commun. Math. Phys. {\bf 79}, 529-572 (1981)

\bibitem[RS78]{RS78} M. Reed, B. Simon, {\it Methods of 
Modern Mathematical Physics IV}, Acad. Press, NY (1978)

\bibitem[LL97]{LL97} E.H. Lieb, M. Loss, {\it Analysis}, Amer.
Math. Society (1997)

\bibitem[JRY00]{JRY00} K. Johnsen, \"O. R\"ognvaldsson, J. Yngvason,
in preparation

\bibitem[A75]{A75} R.A. Adams, {\it Sobolev Spaces}, Acad. Press, NY (1975)



\end{thebibliography}
\end{document}